\documentclass[amsmath,amssymb,prc,final,showpacs,twocolumn]{revtex4} 

\usepackage{bm,hyphenat,xspace} 
\usepackage{graphicx,epsfig}

\newcommand{\scl}{0.63}

\newcommand{\Eqs}{Eqs.}
\newcommand{\Ref}{Ref.}
\newcommand{\Refs}{Refs.}
\newcommand{\mbf}[1]{{\mathbf{#1}}}
\newcommand{\fm}{\;\mathrm{fm}}

\newcommand{\cm}{\mathrm{c\!\:\!.m\!\:\!.}}
\newcommand{\zr}{z_R^{-\frac12}}
\newcommand{\ZaR}{\mathcal{Z}_{\alpha R}^{-\frac12}}
\newcommand{\ZbR}{\mathcal{Z}_{\beta R}^{-\frac12}}
\newcommand{\zR}{\mathcal{Z}}

\begin{document}

\title {Coulomb force effects in low-energy $\alpha$-deuteron scattering}

\author{A.~Deltuva} 
\email{deltuva@cii.fc.ul.pt}
\affiliation{Centro de F\'{\i}sica Nuclear da Universidade de Lisboa, 
P-1649-003 Lisboa, Portugal }

\received{October 10, 2006}


\begin{abstract}
The $\alpha$-proton Coulomb interaction is included in the
description of $\alpha$-deuteron scattering using the
screening and renormalization approach in the framework of
momentum-space three-particle equations. 
The technical reliability of the method is demonstrated.
Large Coulomb-force effects are found.
\end{abstract}

 \maketitle

\section{Introduction \label{sec:intro}}

The application of exact Faddeev three-body theory to the understanding
of nuclear reactions that are dominated, at low energies, by three-body
degrees of freedom has been shadowed in the past by the difficulty in 
dealing with the long-range Coulomb interaction between charged particles.
Up to now this difficulty has been overcome by the use of
approximate theoretical methods, namely, the continuum-discretized 
coupled-channel (CDCC) method \cite{austern:87}, which was proposed
initially to deal with deuteron scattering on a heavier nuclei,
but that is now widely used to analyze data resulting from the reactions
involving halo nuclei.
Given the progress achieved recently in the description of proton-deuteron 
$(pd)$ elastic scattering and breakup using exact three-body equations
\cite{deltuva:05a,deltuva:05d,deltuva:06a}, we are now able to include
the Coulomb interaction in three-body nuclear reactions involving two
charged particles. 
Before we apply the present method to the description of direct nuclear
reactions that are dominated by three-body degrees of freedom,
we show here the results for alpha-deuteron  $(\alpha d)$ scattering which 
at low energies is one of simplest effective three-body nuclear reactions.

The $\alpha d$ scattering has been studied extensively
in the past, both experimentally and theoretically
\cite{gruebler:T,gruebler:79a,bruno:80,koersner:77,slaus:83,niessen:92,%
koike:ad,hahn:85}.
Although $\alpha d$ is a six-nucleon system, at low energies, to a good 
approximation, the $\alpha$ particle may be considered a spin zero 
structureless boson, and thereby the theoretical description of 
$\alpha d$ scattering may be reduced to a  three-body problem made up of one 
$\alpha$ and two nucleons $(N)$. Then, the most serious difficulty
is the treatment of the long-range Coulomb interaction,
which in the previous calculations was either completely neglected
or taken into account only approximately. Using the method developed
in \Refs~\cite{deltuva:05a,deltuva:05d,deltuva:06a} we are now able to include
the Coulomb interaction also for $\alpha d$ scattering, quantitatively
evaluate its importance, and with greater confidence ascertain the quality 
of $\alpha N$  force models.

Section~\ref{sec:th} shortly recalls the technical apparatus 
underlying the calculations.
Section~\ref{sec:res} presents characteristic results.
Section~\ref{sec:concl} gives our conclusions.

\section{Treatment of the Coulomb interaction using the screening
and renormalization approach \label{sec:th} } 

Our treatment of the Coulomb interaction is based on the idea of
screening and renormalization proposed in \Ref~\cite{taylor:74a}
for two-particle scattering and extended in 
\Refs~\cite{alt:78a,alt:02a,deltuva:05a,deltuva:05d} to three-particle
scattering where only two particles are charged.
The Coulomb potential is screened, standard scattering theory is applicable,
and the renormalization procedure is applied to recover the unscreened limit.
The success of the method depends on the choice of 
the screened Coulomb potential 
\begin{gather} \label{eq:wr}
w_R(r) = w(r)\; e^{-(r/R)^n}.
\end{gather}
We prefer to work with a sharper screening than the Yukawa screening
$(n=1)$ of \Ref~\cite{alt:02a}. We want to ensure that the
screened Coulomb potential $w_R(r)$ approximates well the true Coulomb one
$w(r)$ for distances $r$ smaller than the screening radius $R$
and simultaneously vanishes rapidly for $r>R$,
providing a comparatively fast convergence of the partial-wave expansion.
In contrast, the sharp cutoff  $(n \to \infty)$
yields an unpleasant oscillatory behavior in the momentum-space representation,
leading to convergence problems. In \Refs~\cite{deltuva:05a,deltuva:05d}
we found the values $3 \le n \le 6$ to provide a sufficiently smooth,
but at the same time a sufficiently rapid screening around $r=R$;
$n=4$ is our  choice also in the present paper.

We solve Alt-Grassberger-Sandhas (AGS) three-particle scattering
equations~\cite{alt:67a} in momentum space
\begin{subequations}\label{eq:AGS}
  \begin{align} \label{eq:Uba}
     U^{(R)}_{\beta \alpha}(Z) = {} & \bar{\delta}_{\beta \alpha} G_0^{-1}(Z)
     + \sum_{\sigma} \bar{\delta}_{\beta \sigma} T^{(R)}_\sigma (Z) G_0(Z)
     U^{(R)}_{\sigma \alpha}(Z), \\
     U^{(R)}_{0 \alpha}(Z) = {} & G_0^{-1}(Z)
     + \sum_{\sigma}  T^{(R)}_\sigma (Z) G_0(Z) U^{(R)}_{\sigma \alpha}(Z),
  \end{align}
\end{subequations}
with  $\bar{\delta}_{\beta \alpha} = 1 - {\delta}_{\beta \alpha}$,
$ G_0(Z)$ being the free resolvent, 
$T^{(R)}_\sigma (Z)$ the two-particle transition matrix
derived from nuclear plus screened Coulomb potentials, and 
$U^{(R)}_{\beta \alpha}(Z)$  and $U^{(R)}_{0 \alpha}(Z)$ 
the three-particle transition operators
for elastic/rearrangement and breakup scattering; their dependence on the 
screening radius $R$ is notationally indicated.
On-shell matrix elements of the operators \eqref{eq:AGS}
 between  two- and- three-body
channel states  $|\phi_\alpha (\mbf{q}_i) \nu_{\alpha_i} \rangle $
and $|\phi_0 (\mbf{p}_f \mbf{q}_f) \nu_{0_f} \rangle $ with
discrete quantum numbers $\nu_{\sigma_j} $,
Jacobi momenta $\mbf{p}_j $ and  $\mbf{q}_j $,  energy  $E_i$,
and $Z=E_i+i0$, do not have a $R \to \infty$ limit.
However, as demonstrated in \Refs~\cite{alt:78a,deltuva:05a,deltuva:05d},
the three-particle amplitudes can be decomposed into
long-range and Coulomb-distorted short-range parts, where
the quantities diverging in that limit are of two-body nature, i.e., 
the on-shell transition matrix $T^{\cm}_{\alpha R}(Z)$ 
derived from the screened Coulomb potential between
spectator and the center of mass (c.m.) of the bound pair,
the corresponding wave function, and the screened Coulomb wave function
for the relative motion of two charged particles in the final state.
Those quantities, renormalized according to
\Ref~\cite{taylor:74a}, in the $R \to \infty$ limit converge to 
the two-body Coulomb scattering amplitude 
$\langle \phi_\alpha (\mbf{q}_f) \nu_{\alpha_f} |T^{\cm}_{\alpha C}
    |\phi_\alpha (\mbf{q}_i) \nu_{\alpha_i} \rangle$
 (in general, as a distribution)
and to the corresponding Coulomb wave functions, respectively,
thereby yielding the three-particle scattering amplitudes
in the proper Coulomb limit
\begin{subequations}\label{eq:UC}
\begin{gather} \label{eq:UC1}
  \begin{split}
    \langle  \phi_\beta (\mbf{q}_f)  \nu_{\beta_f} & | U_{\beta \alpha}
    |\phi_\alpha (\mbf{q}_i) \nu_{\alpha_i} \rangle   \\ = {}&
    \delta_{\beta \alpha}
    \langle \phi_\alpha (\mbf{q}_f) \nu_{\alpha_f} |T^{\cm}_{\alpha C}
    |\phi_\alpha (\mbf{q}_i) \nu_{\alpha_i} \rangle  \\ & +
    \lim_{R \to \infty} \{ \ZbR(q_f)
    \langle \phi_\beta (\mbf{q}_f) \nu_{\beta_f} |
            [ U^{(R)}_{\beta \alpha}(E_i + i0)  \\ & -
              \delta_{\beta\alpha} T^{\cm}_{\alpha R}(E_i + i0)]
            |\phi_\alpha (\mbf{q}_i) \nu_{\alpha_i} \rangle
            \ZaR(q_i) \},
  \end{split} \\
    \begin{split}
      \langle \phi_0  (\mbf{p}_f \mbf{q}_f) & \nu_{0_f}  | U_{0 \alpha}
      |\phi_\alpha (\mbf{q}_i) \nu_{\alpha_i} \rangle  \\ = {}&
      \lim_{R \to \infty} \{ \zr(p_f)
      \langle \phi_0 (\mbf{p}_f \mbf{q}_f) \nu_{0_f} | \\ & \times
      U^{(R)}_{0 \alpha}(E_i + i0)
      |\phi_\alpha (\mbf{q}_i) \nu_{\alpha_i} \rangle \ZaR(q_i) \}.
    \end{split}
  \end{gather}
\end{subequations}
The renormalization factors $\zR_{\alpha R}(q_j)$ and 
$z_R(p_f)$ are diverging phase 
factors given in \Refs~\cite{taylor:74a,alt:78a,deltuva:05a,deltuva:05d}.
The $R \to \infty$ limit in \Eqs~\eqref{eq:UC}
has to be calculated numerically, but due to the short-range nature
of the corresponding operators it is reached with sufficient accuracy 
at rather modest $R$ if the form of the screened Coulomb potential
has been chosen successfully as discussed above.
More details on the practical implementation of the screening
and renormalization approach are given in
\Refs~\cite{deltuva:05a,deltuva:05d,deltuva:06a}.
Compared to the $pd$ calculations an additional difficulty is the presence of
the $\alpha  N$ $P$-wave transition matrix $T^{(R)}_\sigma (Z)$
poles in the complex energy plane close to the real axis. 
This requires special treatment of those partial waves in the numerical
solution of \Eqs~\eqref{eq:AGS}, 
i.e., the pole factor of $T^{(R)}_\sigma (Z)$ is separated when interpolating
$T^{(R)}_\sigma (Z) G_0(Z) U^{(R)}_{\sigma \alpha}(Z)$ 
and the subtraction technique is used for integration.

\section{Results \label{sec:res}}

There exist in the literature several parametrizations
of the $\alpha N$ potential. Most of them support deeply bound $\alpha N$
state in $S$-wave that is not observed experimentally, since it is forbidden
by the Pauli principle. To account for the Pauli principle that
forbidden state $|b\rangle$ (or some state close to it) has to be projected out
or moved to a large positive energy $\Gamma$, replacing the potential
$v$ by $v' = v + |b\rangle \Gamma \langle b|$. 
As demonstrated in \Ref~\cite{schellingerhout:93a}, the latter method in 
the $\Gamma \to \infty$ limit is equivalent to the first one.  
Alternatively, one may use repulsive $\alpha N$ potential in the $S$-wave.
In order to estimate  the model dependence,
in the calculations of this paper we use three different parametrizations
for the nuclear part of the $\alpha N$ potential, that in the following
will be called $\alpha N$-I, $\alpha N$-II, and $\alpha N$-III.
The potential $\alpha N$-I is taken from \Ref~\cite{thompson:00}.
It is of Woods-Saxon form with central and spin-orbit parts and supports
a bound-state in  $S$-wave that is moved to a large positive energy $\Gamma$;
we found that for $\Gamma \geq 1000$ MeV the results are practically
independent of $\Gamma$.
The potential $\alpha N$-II is taken from \Ref~\cite{thompson:00} as
well and differs from $\alpha N$-I only in $S$-wave where it is repulsive
and local, but, nevertheless, nearly phase equivalent to $\alpha N$-I.
 The third potential $\alpha N$-III, taken from
\Ref~\cite{niessen:92}, is represented as a sum of Gaussians with the strong
repulsive rank-1 separable term in $S$-wave that moves 
the harmonic oscillator ground state which is close to 
the Pauli-forbidden state to $\Gamma = 1000$ MeV.
The potentials fit the low energy experimental $\alpha N$ phase shifts
quite well, although $\alpha N$-III is more attractive in $P$-waves than 
the other two.
All potentials are charge symmetric and act in partial waves with orbital 
angular momentum $L \leq 2$. 
For the neutron-proton $(np)$ interaction we use the charge dependent 
(CD) Bonn potential \cite{machleidt:01a} and include partial waves with 
total angular momentum $I \leq 2$ plus ${}^3D_3$. 
It was checked that those quite low partial waves are 
sufficient for convergence. In contrast, the screened Coulomb potential
is of longer range and therefore, depending on the screening radius $R$,
requires partial waves up to $L \leq 11$. Those high partial waves are
included exactly as described in \Ref~\cite{deltuva:06a}.
We note that the $\alpha p$ Coulomb potential at short distances 
is taken as the one of a uniformly charged sphere, i.e.,
$w(r) = 2\alpha_e [3-(r/R_I)^2]/(2R_I)$ for $r < R_I = 1.5 \fm$,
$\alpha_e$ being the fine structure constant.

\renewcommand{\scl}{0.625}
\begin{figure}[!]
\begin{center}
\includegraphics[scale=\scl]{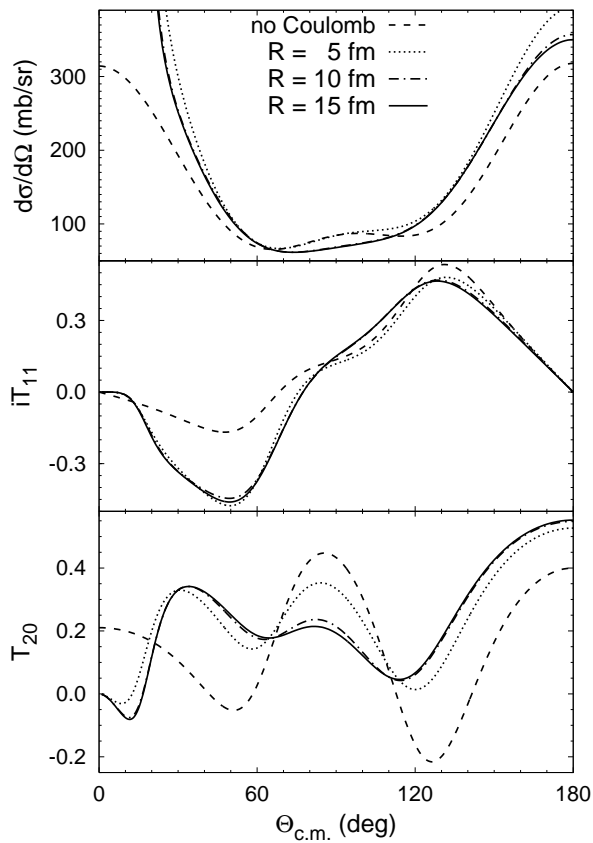}
\end{center}
\caption{\label{fig:Relastic}
Convergence of the $\alpha d$ elastic scattering  observables with 
screening radius $R$. 
The differential cross section and deuteron analyzing powers $iT_{11}$ and
$T_{20}$ at 4.81~MeV deuteron lab energy  are
shown as functions of the c.m. scattering angle.
Results obtained with screening radius $R= 5$~fm (dotted curves),
10~fm (dash-dotted curves), and 15~fm (solid curves) are compared.
Results without Coulomb (dashed curves)
are given as reference for the size of the Coulomb effect.
The calculations were performed with the $\alpha N$-I potential.}
\end{figure}
\begin{figure}[!]
\begin{center}
\includegraphics[scale=\scl]{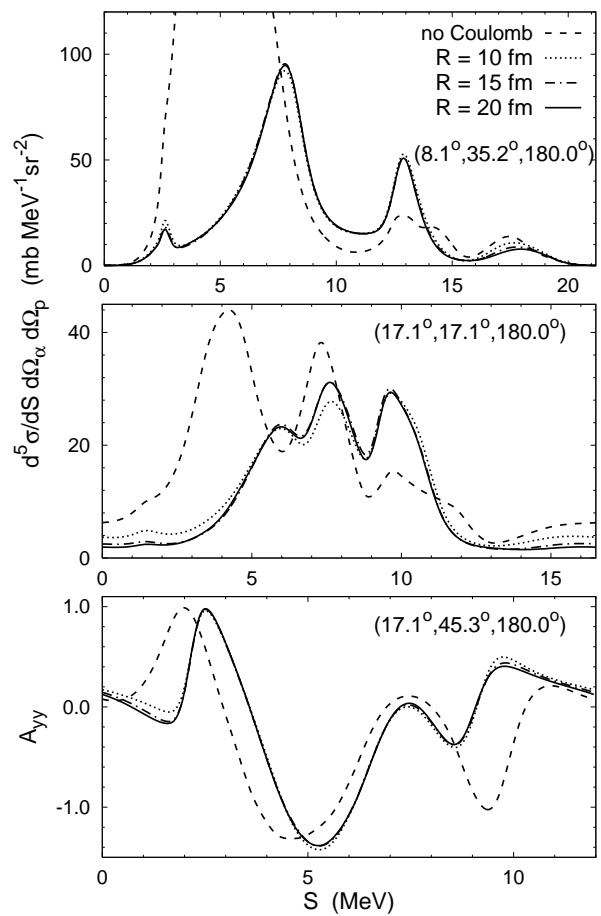}
\end{center}
\caption{\label{fig:Rbreakup}
Convergence of the $\alpha d$ breakup  observables with 
screening radius $R$. 
The differential cross section and deuteron analyzing power $A_{yy}$
in selected kinematical configurations at 15~MeV $\alpha$ lab energy  are
shown as functions of the arclength $S$ along the kinematical curve.
Results obtained with screening radius $R= 10$~fm (dotted curves),
15~fm (dash-dotted curves), and 20~fm (solid curves) are compared.
Results without Coulomb (dashed curves)
are given as reference for the size of the Coulomb effect.
The calculations were performed with the $\alpha N$-I potential.}
\end{figure}

The internal criterion for the reliability of the screening and 
renormalization approach is the convergence
of the observables with the screening radius $R$ employed to calculate the
Coulomb-distorted short-range part of the amplitudes in \Eqs~\eqref{eq:UC};
that criterion was found to be absolutely reliable for $pd$ scattering
\cite{deltuva:05b}. Figures \ref{fig:Relastic} and \ref{fig:Rbreakup}
show several examples for $\alpha d$ elastic scattering and breakup.
The breakup kinematical final-state configurations are characterized
in a standard way by the polar angles of the $\alpha$-particle and the proton
and by the azimuthal angle between them,
$(\theta_\alpha,\theta_p,\varphi_{p}-\varphi_{\alpha})$.
The definition of the arclength $S$ along the kinematical curve is the
standard one as in Refs.~\cite{koersner:77,slaus:83,niessen:92}.
Even when the Coulomb effect is large
the convergence is impressively fast, e.g.,
the screening radius $R = 10 \fm$ is practically sufficient for the elastic
$\alpha d$ scattering at 4.81~MeV deuteron lab energy in 
Fig.~\ref{fig:Relastic}, while the observables of $\alpha d$ breakup
at 15~MeV $\alpha$ lab energy in Fig.~\ref{fig:Rbreakup}
require at least $R=15 \fm$ for convergence. As discussed in 
\Refs~\cite{deltuva:05a,deltuva:05d}, the convergence rate is energy 
dependent, i.e., larger screening radii are needed for  elastic
scattering observables at very low energies and for the breakup
differential cross section in kinematical situations characterized 
by very low  relative energy of the two charged particles, i.e.,
close to the $\alpha p$ final-state interaction (FSI) regime.
Nevertheless, the observed convergence strongly suggests the reliability
of the present Coulomb treatment using screening and renormalization
approach.

As already shown in Figs.~\ref{fig:Relastic} and \ref{fig:Rbreakup}
the Coulomb effect on the observables of low energy $\alpha d$ scattering
may be very strong, indicating that the inclusion of the Coulomb interaction
is necessary for a stringent comparison of the theoretical results and 
experimental data.
Obviously, we have many more predictions than it is possible to show.
Therefore we make a selection of the most interesting predictions
which illustrate the message we believe the results tell us.
The readers are welcome to obtain the results for their favorite data from us.

\renewcommand{\scl}{0.65}
\begin{figure*}[!]
\begin{center}
\includegraphics[scale=\scl]{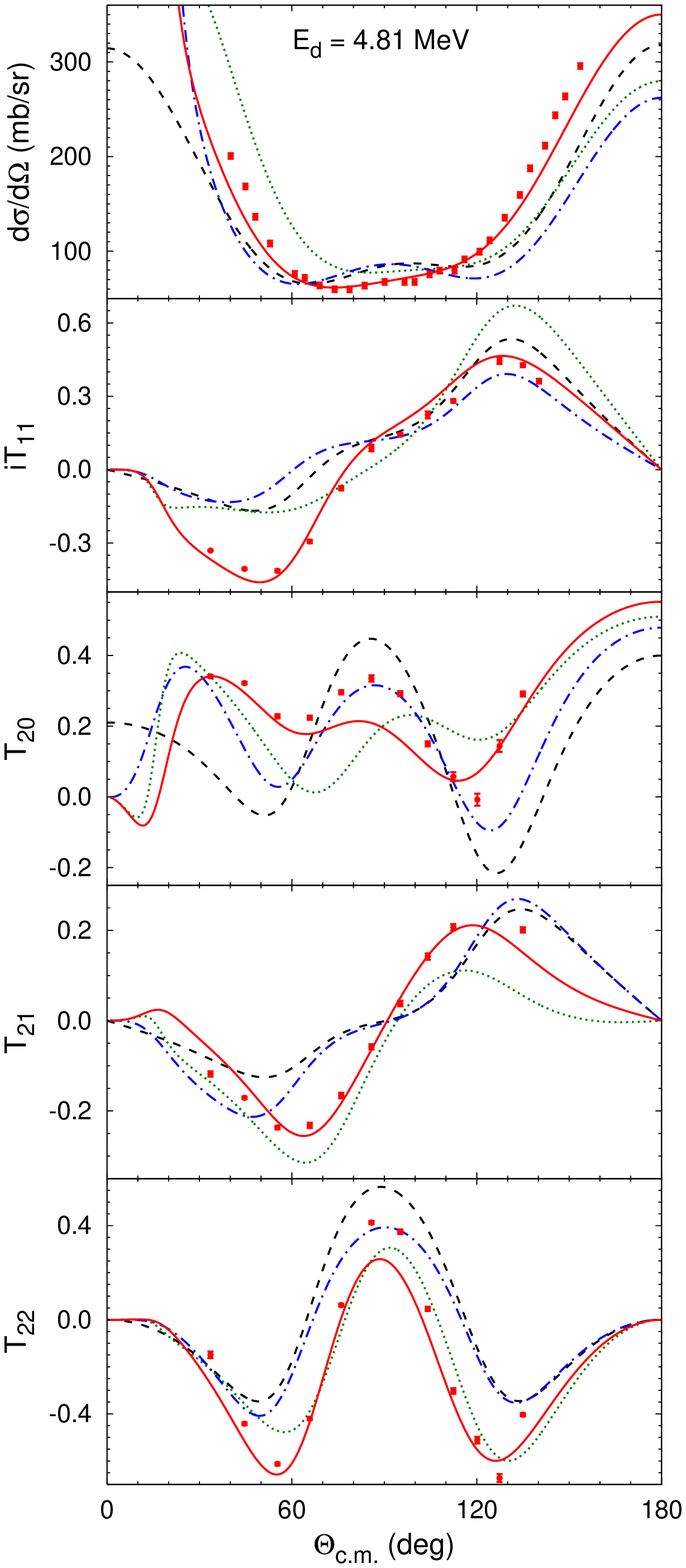}
\includegraphics[scale=\scl]{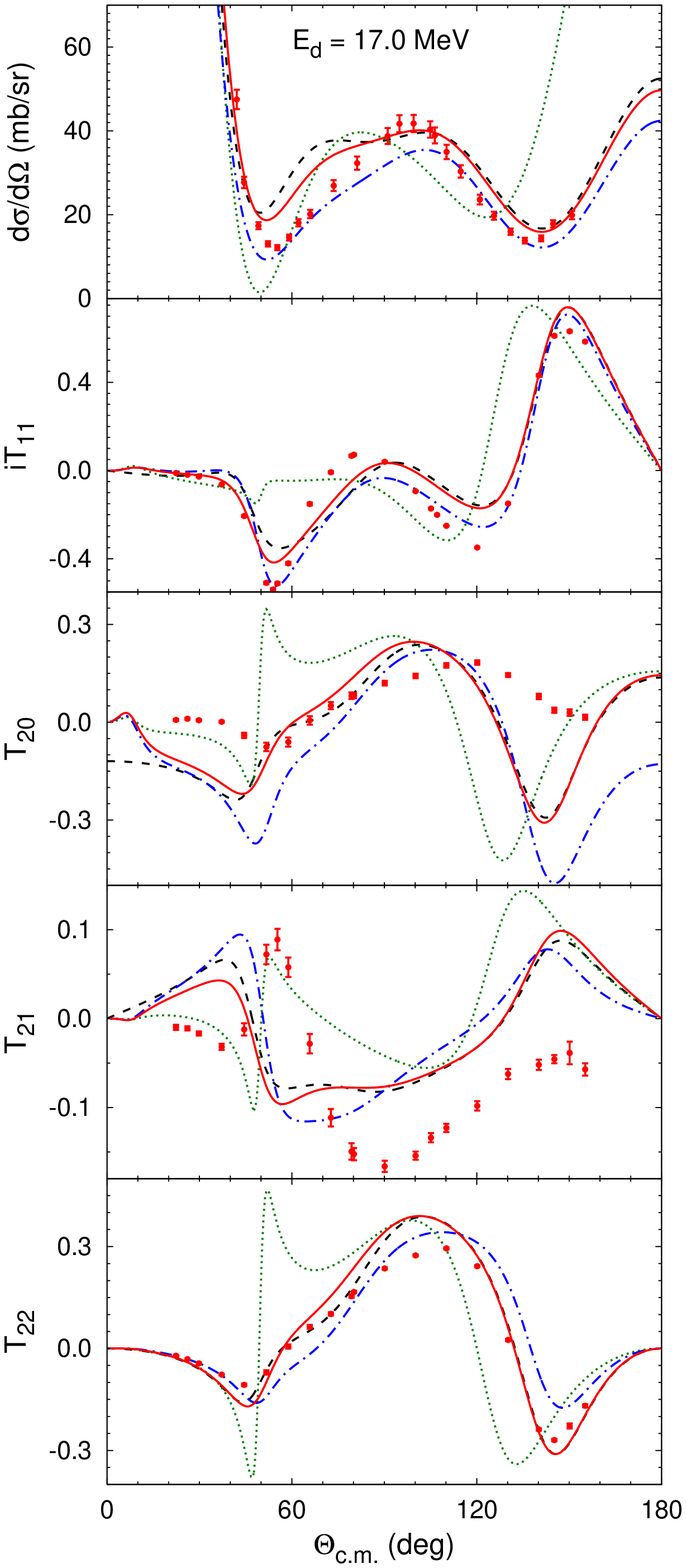}
\end{center}
\caption{\label{fig:elastic} (Color online) 
The differential cross section and deuteron analyzing powers 
at 4.81~MeV and 17~MeV deuteron lab energy.
Results including the Coulomb interactions obtained with potentials
$\alpha N$-I (solid curves), $\alpha N$-II (dotted curves),
and $\alpha N$-III (dash-dotted curves) are compared.
$\alpha N$-I results without Coulomb (dashed curves)
are given as reference for the size of the Coulomb effect.
The experimental 4.81-MeV data  are from 
\Ref~\cite{bruno:80} for the differential cross section and from
\Ref~\cite{gruebler:T} for the spin observables,
and  the 17-MeV data are from \Ref~\cite{gruebler:79a}.}
\end{figure*}

Figure~\ref{fig:elastic} presents our results for the differential cross
section and all deuteron analyzing powers of elastic $\alpha d$ scattering
at 4.81 MeV and 17 MeV deuteron lab energy. As expected, the Coulomb effect 
is large at lower energy for all observables in the whole angular regime,
while with increasing energy it becomes smaller.
Furthermore, the results depend strongly on the choice of the 
$\alpha N$ interaction. The potential $\alpha N$-I describes most of
the experimental $E_d = 4.81$ MeV data quite satisfactorily, 
whereas the potential $\alpha N$-II that avoids the Pauli forbidden state 
in $S$-wave by the local repulsion clearly fails in accounting for the 
scattering data, especially at $E_d = 17$ MeV,
although the ${}^6\mathrm{Li}$ bound state properties predicted by
both potentials are very similar.
The potential $\alpha N$-III also fails in accounting for the experimental 
data at $E_d = 4.81$ MeV, but the difference between predictions using 
$\alpha N$-I and $\alpha N$-III is found to be mostly due to the 
differences in $P$-waves. At $E_d = 17$ MeV, 
which is just 3 MeV below ${}^3\mathrm{H}+{}^3\mathrm{He}$ 
threshold, there is a qualitative agreement between experimental data and 
predictions with $\alpha N$-I and $\alpha N$-III,
except for deuteron tensor analyzing powers $T_{20}$ and $T_{21}$.
Obviously, the treatment of the $\alpha$-particle as a structureless boson 
becomes less reliable with increasing energy.

\renewcommand{\scl}{0.65}
\begin{figure*}[!]
\begin{center}
\includegraphics[scale=\scl]{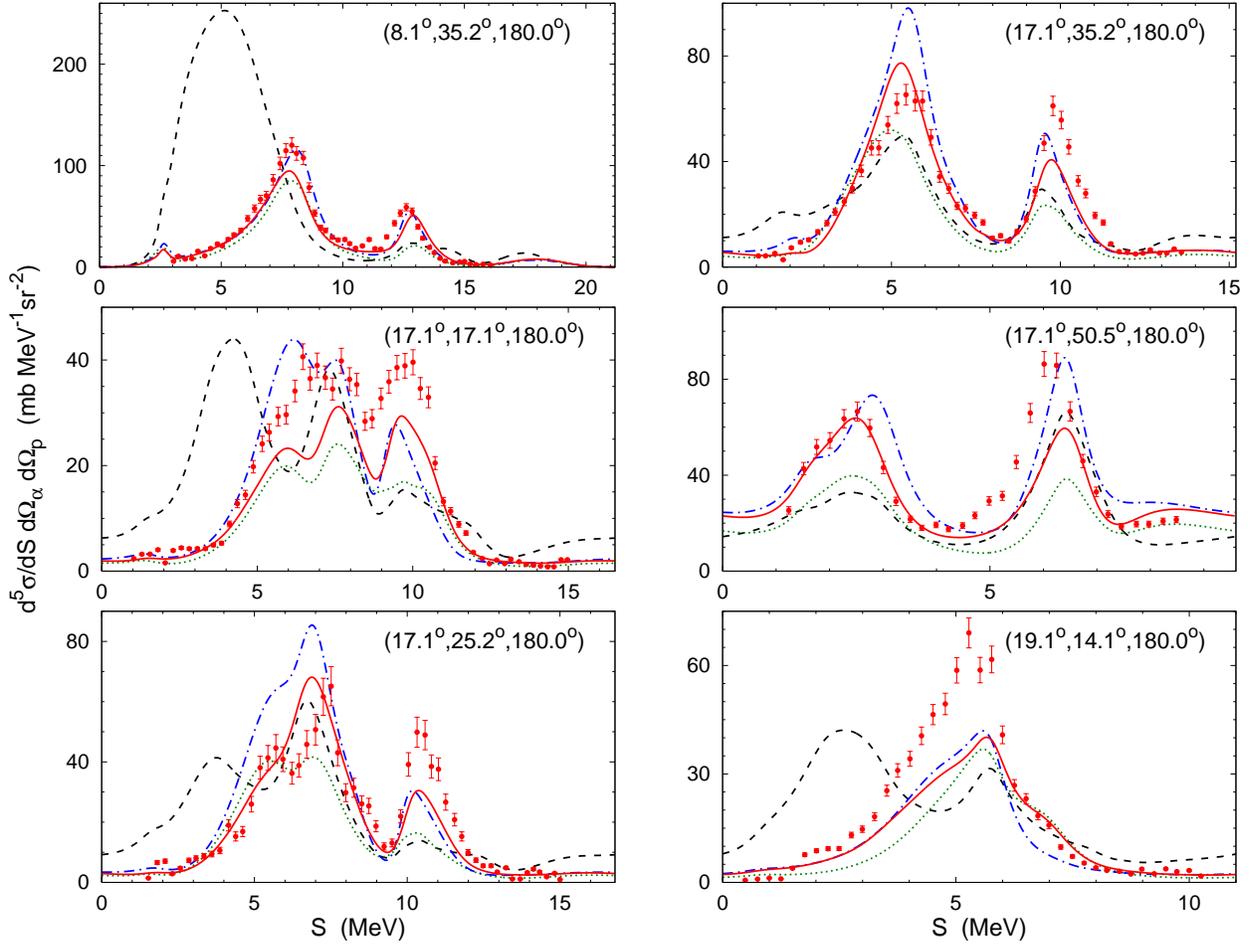}
\end{center}
\caption{\label{fig:breakup} (Color online) 
The differential cross section of the $\alpha d$ breakup
at 15~MeV $\alpha$ lab energy in selected kinematical configurations.
Curves as in Fig.~\ref{fig:elastic}.
The experimental data are from \Ref~\cite{koersner:77}.}
\end{figure*}
\begin{figure*}[!]
\begin{center}
\includegraphics[scale=\scl]{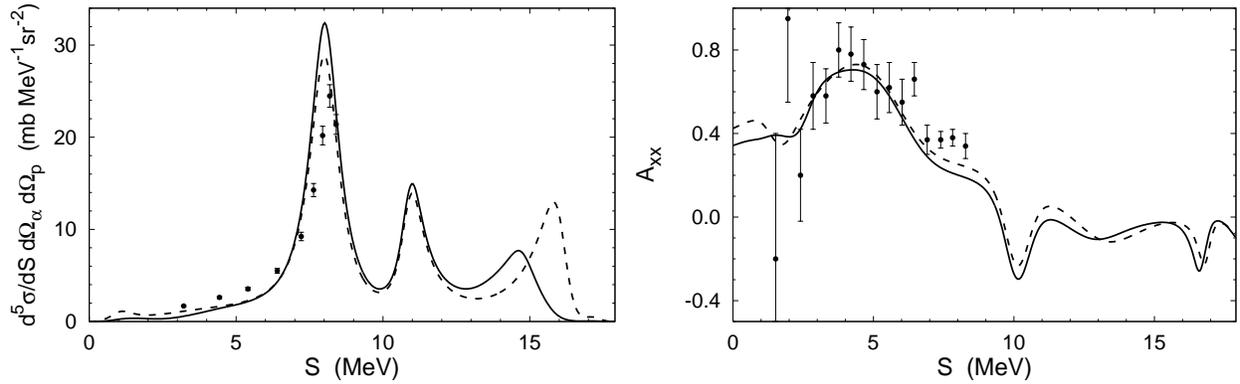}
\end{center}
\caption{\label{fig:breakup12}
The differential cross section and deuteron analyzing power $A_{xx}$
of the $\alpha d$ breakup at 12~MeV deuteron lab energy
in the $(25^{\circ},45^{\circ},180^{\circ})$ configuration.
Results obtained using the potential $\alpha N$-I  with (without)
Coulomb are shown as solid (dashed) curves. 
The experimental data are from \Ref~\cite{slaus:83}.}
\end{figure*}

Figure \ref{fig:breakup} presents our results for the differential cross 
section of $\alpha d$ breakup at 15~MeV $\alpha$ lab energy
in various kinematical configurations. 
The most important Coulomb effect is the shift of
the  $\alpha p$ $P$-wave resonance position that leads to 
the corresponding changes in the structure of the observables.
Furthermore, the Coulomb interaction breaks $\alpha n - \alpha p$
charge symmetry and thereby allows the coupling to the $np$ isospin triplet 
waves, in particular ${}^1S_0$.
The predictions without Coulomb fail completely in accounting for the 
experimental data, while inclusion of the Coulomb moves the peaks of
the differential cross section to the right positions, although the
height of those peaks is not always reproduced. 
There is a strong dependence on the employed  $\alpha N$ potential,
where, like in the elastic scattering,
the potential $\alpha N$-II provides the worst description of the data,
and the difference between the predictions of
$\alpha N$-I and $\alpha N$-III is dominated by the $P$-waves.
The sensitivity of the results to the choice of a realistic $np$ potential
is also checked and found to be insignificant.
When the c.m. energy increases one finds a larger part of the phase space
where the relative  $\alpha p$ energy may be quite different from the one 
corresponding to the $P$-wave resonance and therefore also the Coulomb effect
is less significant as shown in Fig.~\ref{fig:breakup12}.

\section{Summary \label{sec:concl}}

In this paper we show that the screening and renormalization approach
for the inclusion of the Coulomb interaction in the description of
three-particle scattering using momentum-space integral equations,
developed in \Refs~\cite{deltuva:05a,deltuva:05d,deltuva:06a} for $pd$ 
reactions, can be extended reliably to $\alpha d$ scattering.
This is an important step towards application of exact scattering
equations for the description of nuclear reactions within three-body models,
where up to now only approximate treatments like the 
CDCC method \cite{austern:87} have been used.

The Coulomb effect on the observables of $\alpha d$ scattering is studied
and is found to be large in elastic scattering at very low energies
and in breakup, where the shift of $\alpha p$ $P$-wave resonance
position leads to the corresponding shifts of the differential cross 
section peaks.

Another important consequence of this work is that it allows to
ascertain with greater confidence the quality of the $\alpha N$
force models one uses to describe $\alpha d$ observables 
as well as structure and reactions of ${}^6\mathrm{Li}$ and ${}^6\mathrm{He}$.
Although at present there are too large uncertainties in the parametrization
of $P$- and $D$-wave interactions that need to be improved, 
this work clearly indicates the superiority of the attractive 
$S$-wave potentials supporting a Pauli-forbidden state (that is
projected out or moved to a large positive energy) over the repulsive
$S$-wave potentials.

\vspace{2mm}
\begin{acknowledgments}
The author thanks A.~C.~Fonseca for many discussions and suggestions,
and  W.~Gr\"uebler for providing experimental data.
The work is supported by the Funda\c{c}\~{a}o para a Ci\^{e}ncia e a Tecnologia
(FCT) grant SFRH/BPD/14801/2003.
\end{acknowledgments}


\bibliographystyle{prsty}

\end{document}